\documentstyle[eprocl,12pt,psfig]{article}
\newcommand{\bucky}{C$_{60}$\ }
\begin{document}
\title{QUANTUM MONTE-CARLO CALCULATIONS FOR INTEGER-DOPED FULLERIDES}
\author{ERIK KOCH$^{a}$, OLLE GUNNARSSON$^{b}$, and RICHARD MARTIN$^{a}$} 
\address{
  $^{a)}$ Department of Physics and Materials Research Laboratory,\\ 
          University of Illinois, Urbana, IL 61801, USA\\
  $^{b)}$ Max-Planck-Institut f\"ur Festk\"orperforschung, 
         70569 Stuttgart, Germany}
\maketitle
\abstracts{
The doped Fullerides can be well described by a Hubbard model, which comprises
the partly filled, threefold-degenerate $t_{1u}$ orbital and the on-site Coulomb
interaction $U$. The orbital degeneracy is known to shift the critical ratio
$U_c/W$ for the Mott-Hubbard transition towards larger values. This puts the
half-filled alkali-doped Fullerides A$_3$\bucky on the metallic side of the
transition. Prompted by the recent synthesis of isostructural families of 
integer-doped Fullerides with different fillings, we investigate how the orbital
degeneracy affects the Mott transition at integer fillings $n\ne3$. The 
calculations are done by fixed-node Diffusion Monte-Carlo, using a trial
function, which permits us to systematically vary the magnetic character of the
system.}
  
\section{Introduction}
Due to the relatively large distance between the \bucky molecules in solid 
Fullerides
the molecular energy levels merely broaden into narrow subbands. Doping the
solid with alkali atoms has the effect of filling the subband derived from the
LUMO, the $t_{1u}$ orbital. Band structure calculations indicate that the 
electronic structure changes little upon doping, if the compounds retain the
same crystal structure.\cite{Erwin} The width of the $t_{1u}$ band should thus
be fairly independent of doping, and, for the Fullerides with fcc structure 
exhibiting orientational disorder, is estimated to be $W\approx 0.5\ldots
0.8\;eV$.\cite{c60mott} The Coulomb interaction between two electrons on the 
same \bucky molecule is in the range $U\approx 1.2\ldots 1.5\;eV$, and should 
also be quite independent of doping.\cite{HubbardU} 

An effective Hamiltonian for the electrons in the partially filled band is
given by a multi-band Hubbard model, taking into account the hopping between
the $t_{1u}$-orbitals on neighboring molecules and the 
on-site Coulomb interaction $U$. It turns out that, as a consequence of the 
three-fold degeneracy of the $t_{1u}$-orbital, A$_3$\bucky is a strongly 
correlated metal, close to a Mott-transition.\cite{c60mott} 
Here we investigate the 
critical ratio $U_c/W$ at which the metal-insulator transition takes place
for the other, integer-doped Fullerides A$_n$\bucky with fcc structure.
Our work has been prompted by the recent synthesis of isostructural
families of doped Fullerides.\cite{Yildirim} We give an intuitive argument for 
the doping-dependence of the Mott-transition, and present results of Quantum
Monte-Carlo calculations indicating that the metal-insulator transition is
shifted towards smaller $U_c/W$
as one moves away from half-filling.

\section{Degeneracy Enhancement}

\begin{figure}[b]
\centerline{\psfig{figure=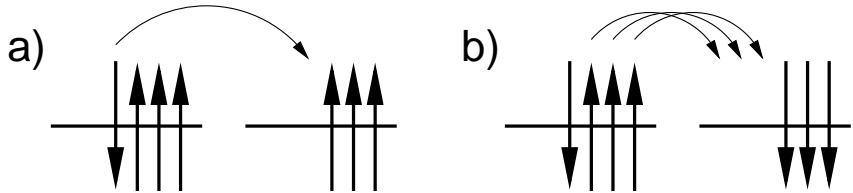,height=1.0in}}
\caption{\label{hopping}
 Illustration of how an extra electron can hop against an integer-filled
 background (here: degeneracy $N=3$, filling $n=3$). 
 a) ferromagnetically aligned neighbor: Only one electron can hop.
 b) antiferromagnetically aligned neighbor: There are three hopping channels. 
}
\end{figure}

We first consider the effect of orbital degeneracy on the Mott-Hubbard
transition in a highly idealized model. A system with $L$ electrons becomes
insulating when the conduction gap
\begin{equation}\label{Egap}
 E_g=E(L+1)-2\,E(L)+E(L-1)
\end{equation} 
opens. In the limit of large Coulomb interaction $U\gg t$, we can estimate 
the energies entering eqn.~(\ref{Egap}). For integer filling $L=n\,M$, where M 
is the number of molecules in the system, all molecules will be occupied by
$n$ valence electrons. Hopping is strongly suppressed, since it increases the
energy by $U$. Hence for the integer filled system, to leading order in 
$t^2/U$, there is no kinetic contribution to the total energy 
\begin{equation}
 E(L)={n(n-1)\over2}\,M\;U + {\cal O}\Big(t^2/U\Big)
\end{equation}
In contrast, the systems with $L=n\,M\pm 1$ electrons have an extra 
electron/hole that can hop without an additional cost in Coulomb energy. To 
estimate the kinetic contribution to the total energy, we calculate the hopping 
matrix-element for hopping of the extra charge to a neighboring molecule. We 
notice that for the non-degenerate Hubbard model a ferromagnetic arrangement of
the spins is energetically favored (Nagaoka's theorem for bipartite 
lattices \cite{Nagaoka}). For a
degenerate Hubbard model, however, the hopping matrix element is larger for
an antiferromagnetic alignment of the spins on neighboring sites. In that case,
as illustrated in Fig.~\ref{hopping}, instead of only one, several equivalent 
electrons can hop. To estimate the matrix element for nearest-neighbor hopping,
we calculate the matrix element for hopping of the extra electron/hole against
a N\'eel state $|0\rangle$. 
Denoting the state with the extra charge on site $i$
by $|i\rangle$, we find that the second moment of the Hamiltonian 
$\langle i|H^2|i\rangle$ is given by the number of hopping channels $k$ times 
the number of (equivalent) nearest-neighbors $z$ times the single-electron 
hopping matrix-element $t$ squared. By inserting the identity in the form 
$\sum_j\;|j\rangle\langle j|$, where $|j\rangle$ denotes the state with the 
extra charge hopped from site $i$ to the neighboring site $j$, we then find
\begin{equation}
  \langle i|H|j\rangle = \sqrt{k}\;t
\end{equation}
i.e.\ {\em the hopping matrix-element is enhanced by a factor of $\sqrt{k}$ 
compared to the one-particle case.} Taking the analogy with the one-particle
problem further, we then expect the energy of the system to be lowered by 
$\sqrt{k}$ times half the single-electron band-width $W$.
The gap in the limit $U\gg t$ is thus given by
\begin{equation}
  E_g = U - \Big(\sqrt{k_{L+1}} + \sqrt{k_{L-1}}\,\Big)\;W/2 .
\end{equation}
Without degeneracy we have $E_g=U-W$, while for orbital degeneracy $N=3$ we 
find
\begin{displaymath}
\begin{array}{c|c|c|c}
\mbox{filling } & n=1,\;5 & n=2,\;4 & n=3 \\ \hline
\raisebox{0pt}[3.01ex][0ex]{}
(\sqrt{k_{L+1}}+\sqrt{k_{L-1}})/2  
& (\sqrt{2}+1)/2        \approx 1.21
& (\sqrt{3}+\sqrt{2})/2 \approx 1.57
&  \sqrt{3}             \approx 1.73
\end{array}
\end{displaymath}
i.e.\ extrapolating the result to intermediate $U$, we expect that the degeneracy enhancement shifts the critical ratio
$U_c/W$ to larger values, compared to the non-degenerate case. The shift is
largest for half-filling and becomes smaller as one moves away from $n=3$. 
The symmetry of the result around half-filling reflects the electron-hole
symmetry (we have implicitly assumed a bipartite lattice). We stress that
the above argument is not rigorous, its main purpose is to give an idea of
how orbital-degeneracy affects the electron dynamics.

\section{Quantum Monte-Carlo Results}
To investigate the doping dependence of the Mott-transition in the doped 
Fullerides, we have performed Quantum Monte-Carlo calculations for fcc clusters
of up to 32 molecules, using a multi-band Hubbard-Hamiltonian. The hopping integrals have 
been obtained from a tight-binding parameterization,\cite{TBparam} while $U$ 
has been varied 
to find the Mott-transition. The ground state energies entering in 
eqn.~(\ref{Egap}) have been determined by diffusion Monte-Carlo calculations 
using the fixed-node approximation.\cite{tenHaaf} As trial function we have 
used the product of a Slater-determinant and a Gutzwiller factor. The 
Slater-determinant has been determined from a Hartree-Fock (HF) calculation. 
To allow for different magnetic characters of the trial function, we have 
treated the Coulomb-interaction $U_0$ in the HF calculation as a variational 
parameter.

\begin{figure}
\centerline{\psfig{figure=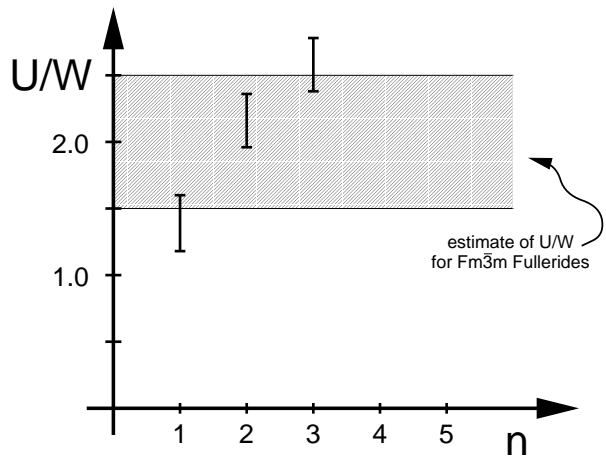,height=2.4in}}
\vspace{-1.5ex}
\caption{\label{QMCUc}
 Estimate of the critical ratio $U_c/W$ for a multi-band Hubbard model 
 describing the doped Fm${\bar 3}$m Fullerides at integer fillings $n$. 
 The bars give our estimates for $U_c/W$. The hatched region indicates 
 the $U/W$ range in which these Fullerides are believed to lie.
}
\end{figure}

Our results for fillings $n=1\ldots3$ are shown in Fig.~\ref{QMCUc}. As 
suggested by the degeneracy argument, we find that the critical ratio decreases
as we move away from half-filling. Under the caveat that the neglected effects 
(higher subbands, multiplet effects, electron-phonon coupling) do not change the
results drastically, we can conclude that the hypothetical doped Fullerides at
low temperatures {\em with Fm${\bar 3}$m structure} are strongly correlated 
systems with A$_2$\bucky being very close to the Mott transition and 
$A_1$\bucky possibly already being in the insulating regime.

\section*{Acknowledgments}
This work has been supported by the Department of Energy under grant 
DEFG 02-91ER45539. The computations were performed on the SP2 at the Cornell 
Theory Center. One of the authors (E.K.) wants to thank the 
Alexander-von-Humboldt-Stiftung for financial support under the 
Feodor-Lynen-Program.

\section*{References}
\def\Journal#1#2#3#4{{#1} {\bf #2}, #3 (#4)}
\def\PRL{\em Phys.~Rev.~Lett.}
\def\PRB{{\em Phys.~Rev.}~B}

\end{document}